\documentclass[journal,draftcls,onecolumn,12pt,twoside]{IEEEtran}

\usepackage{amsmath,graphicx}
\usepackage{amsfonts}
\usepackage{bm}
\usepackage{amsthm}
\newtheorem{proposition}{Proposition}
\usepackage{color}
\usepackage{multirow}
\newcommand{\tred}{\color{red}}
\begin{document}
%
% paper title
% Titles are generally capitalized except for words such as a, an, and, as,
% at, but, by, for, in, nor, of, on, or, the, to and up, which are usually
% not capitalized unless they are the first or last word of the title.
% Linebreaks \\ can be used within to get better formatting as desired.
% Do not put math or special symbols in the title.
\title{Statistical Analysis of a Posteriori Channel and Noise Distribution
Based on HARQ Feedback}
%
%
% author names and IEEE memberships
% note positions of commas and nonbreaking spaces ( ~ ) LaTeX will not break
% a structure at a ~ so this keeps an author's name from being broken across
% two lines.
% use \thanks{} to gain access to the first footnote area
% a separate \thanks must be used for each paragraph as LaTeX2e's \thanks
% was not built to handle multiple paragraphs
%

\author{
  Wenhao Wu,~\IEEEmembership{Student Member,~IEEE,}
  Hans Mittelmann,
  Zhi Ding,~\IEEEmembership{Fellow,~IEEE}
}

% make the title area
\maketitle

% As a general rule, do not put math, special symbols or citations
% in the abstract or keywords.
\begin{abstract}
  In response to a comment on one of our manuscript, this work studies the
  posterior channel and noise distributions conditioned on the NACKs and ACKs of
  all previous transmissions in HARQ system with statistical approaches. Our main
  result is that, unless the coherence interval (time or frequency) is large as
  in block-fading assumption, the posterior distribution of the channel and
  noise either remains almost identical to the prior distribution, or it mostly
  follows the same class of distribution as the prior one. In the latter case,
  the difference between the posterior and prior distribution can be modeled as
  some parameter mismatch, which has little impact on certain type of
  applications.
\end{abstract}

% Note that keywords are not normally used for peerreview papers.

% For peer review papers, you can put extra information on the cover
% page as needed:
% \ifCLASSOPTIONpeerreview
% \begin{center} \bfseries EDICS Category: 3-BBND \end{center}
% \fi
%
% For peerreview papers, this IEEEtran command inserts a page break and
% creates the second title. It will be ignored for other modes.
\IEEEpeerreviewmaketitle

\section{Introduction}
\label{sec:intro}
In our recent manuscript~\cite{wu2015modulation}, we studied the Modulation
Diversity (MoDiv) design problem based on Hybrid Automatic Repeat reQuest (HARQ)
Chase-Combining (CC) protocol. In this work, we approximate the bit
error rate (BER) based on prior fading channel and noise distributions after
each round of (re)transmission. One would argue that the need for the
$m$-th (re)transmission implies that all previous transmissions have failed,
thus the posterior channel and noise distribution would no longer be the same as
the prior distribution. Consequently, it is natural to wonder when and how the
posterior information, namely HARQ NACKs/ACKs, affects the channel and noise
distribution.

There are a few works highlighting this difference between the prior and
the posterior
distributions~\cite{gu2006modeling}\cite{long2012analysis}\cite{alkurd2015modeling},
suggesting that adopting the prior distribution may lead to an overoptimistic
estimation on the performance of HARQ.
On the other hand, there are also abundant works about HARQ in fading channels
that do not consider the posterior distribution, such as constellation
rearrangement~\cite{harvind2005symbol}, power
allocation~\cite{chaitanya2014adaptive}, rate selection~\cite{jin2011optimal}
and so forth. As far as we know, it remains an open question
under what conditions it is suitable or not to exploit the posterior distribution.

In this work, we study the posterior distribution of fading channels and noises
in a practical LDPC-coded HARQ system under the general a priori assumptions of
Rician fading channel and circularly symmetric complex Gaussian (CSCG)
additive noises. By analyzing the posterior distribution with a series of three
hypothesis tests on numerically generated channel and noise samples, we
demonstrate that the posterior distribution may not significantly differ from
the prior one, especially when each HARQ packet, or transport block (TB) in LTE
terminology, experiences a few independent fading channel instances. 
Moreover, even when the coherence interval is large so that the instances of
fading channels corresponding to each TB are more correlated, the posterior
distribution may still follow the same type of distribution as the prior one
except for some differences in parameters. This minor difference has negligible
impact on specific applications such as modulation diversity (MoDiv)
design~\cite{wu2015modulation}\cite{harvind2005symbol}. To the best of our
knowledge, the statistical approaches taken by this work to study the posterior fading channel and noise
distribution in HARQ systems has not been reported in existing literature.

The rest of the paper is organized as follows:
Section~\ref{sec:backgrounds} discusses a few
practical considerations why it may not be proper to adopt the posterior
distribution in the studies of HARQ system. Section~\ref{sec:model_data}
describes our system model and how we generate the fading channel/noise samples
corresponding to the posterior distribution for our hypothesis tests.
In Section~\ref{sec:tests}, we construct three hypothesis tests to analyze the
posterior fading channel/noise distribution. The numerical results are provided
in Section~\ref{sec:numerical}. Finally, Section~\ref{sec:conclusion} concludes
this work.

% needed in second column of first page if using \IEEEpubid
%\IEEEpubidadjcol

\section{Backgrounds}
\label{sec:backgrounds}

One apparent reason why posterior distribution is not preferable is
infeasibility, as a posterior analysis for HARQ system is usually too difficult
unless one rely on some very restrictive, less practical settings and assumptions. For instance, \cite{gu2006modeling} characterize the failure of
transmissions with effective SNR and rate criteria, which on its own is a
simplification and only numerical results are presented. Considering PAM
constellations and maximum ratio combining (MRC),
\cite{long2012analysis}\cite{alkurd2015modeling} attempts to explicitly
formulate the error probability. However, the error probability based on $Q$
function is an approximation, especially for practical high-order QAM
modulations, and their analysis is based on instantaneous CSI and does not scale
well for large number of retransmissions. In practice, a transmission failure is
declared by the cyclic redundancy check (CRC) when an error in the forward error
correction (FEC) decoding result is detected. Such a complex event is difficult
to characterize, let alone deriving a posterior channel distribution from it.

There is one questionable assumption common to all these works considering the
posterior distribution in HARQ.
The channel corresponding to a TB is always characterized by a single scalar
effective SNR value, i.e. the entire TB experience a single instance of the
fading channel and/or additive noise. As a practical example, in LTE system,
each TB can be mapped to a maximum of 110 resource blocks (RB) of
0.5ms$\times$180kHz~\cite[Table 7.1.7.2.1-1]{ts36.213}.
In the propagation condition~\cite[Table B.2-3]{ts36.141}, the coherence time
could be as small as $\tau_c\approx 1 / (4 \times 300\mbox{Hz}) =
0.833\mbox{ms}$ and the coherence bandwidth could be $B_c\approx 1 / (2 \times
5000\mbox{ns}) = 100\mbox{kHz}$~\cite[Table 2.1]{tse2005fundamentals}.
Consequently, each RB roughly experiences independent fading components and the
univariate fading/noise instance per TB assumption is not satisfied. On the
other hand, if each TB experiences $N_{IF}$ independent fading channel
instances, then the posterior channel/noise distribution should be defined over
$\mathcal{O}(m\times N_{IF})$ complex variables, which easily becomes
intractable.

The generalization from the abovementioned restrictive settings and assumptions 
which facilitates a posterior analysis for HARQ leads to the
second---more essential but less obvious---reason why posterior distribution is
not always worthy of exploiting: questionable necessity. In the rest part of
this work, we will demonstrate that, in a more general and practical HARQ system, the
posterior channel and noise distributions may not differ significantly from the
prior ones, or the difference is too little to have visible
impact on certain applications.

\section{System Model and Data Generation}
\label{sec:model_data}

\subsection{Notations}
\label{subsec:notation}
We adopt the following notations throughout this work. $\Re\{\cdot\}$ and
$\Im\{\cdot\}$ represent the real and imaginary part of a complex matrix.
$[\mathbf{A};\mathbf{B}]$ and $[\mathbf{A},\mathbf{B}]$ represent vertical and
horizontal concatenation of matrix $\mathbf{A}$ and $\mathbf{B}$, respectively. Multivariate Gaussian distribution, 
multivariate CSCG distribution and chi-squared distribution with $d$
degree-of-freedom are denoted with $\mathcal{N}(\cdot)$, $\mathcal{CN}(\cdot)$
and $\chi_d^2$. $\mathbf{0}_l$, $\mathbf{1}_l$ and $\mathbf{I}_l$ denote the
$l$-dimensional all-0 vectors, $l$-dimensional all-1 vectors and
$l$-by-$l$-dimensional identity matrix. $|\cdot|$ and $\|\cdot\|_F$ represent
the deterministic and Frobenius norm of a matrix. $\mbox{diag}(\mathbf{a})$
represent the diagonal matrix whose diagonal elements are defined by vector
$\mathbf{a}$.

\subsection{System Model}
\label{subsec:model}
We consider a simple Type-I HARQ system with Chase Combining (CC) under Rician
fading channel and additive CSCG noise assumption. The received signal of the
$m$-th retransmission ($m=0$ represents the original transmission) is
\begin{equation}
  y^{(m)} = h^{(m)}s^{(m)} + n^{(m)}, \quad
\end{equation}
where $s^{(m)}$ is the transmitted symbols originated from the same bit sequence
across the (re)transmissions, and $n^{(m)}\sim\mathcal{CN}(0,\sigma^2)$ is the
additive noise.
The Rician channel can be modeled as~\cite[(2.55)]{tse2005fundamentals}
\begin{align}
  h^{(m)} = \sqrt{\frac{K}{K+1}\beta}e^{j\theta} +
  \sqrt{\frac{1}{K+1}}\mathcal{CN}(0,\beta)
\end{align}
where $K$ is the Rician factor, $\beta$ is the mean power, and $\theta$ is the
phase of the line-of-sight (LOS) component. We also assume that $n^{(m)}$ is
independent across different samples and $h^{(m)}$ is independent across different
(re)transmissions. Assuming all previous $m$ decoding attempt have failed,
after the $m$-th retransmission, the receiver makes another decoding attempt by
combining the $m + 1$ TBs received so far using a maximum likelihood (ML)
detector, until $m > M$ where the HARQ transmission fails.

\subsection{Data Generation}
\begin{figure}[!t]
  \centering
  \includegraphics[width=4.0in]{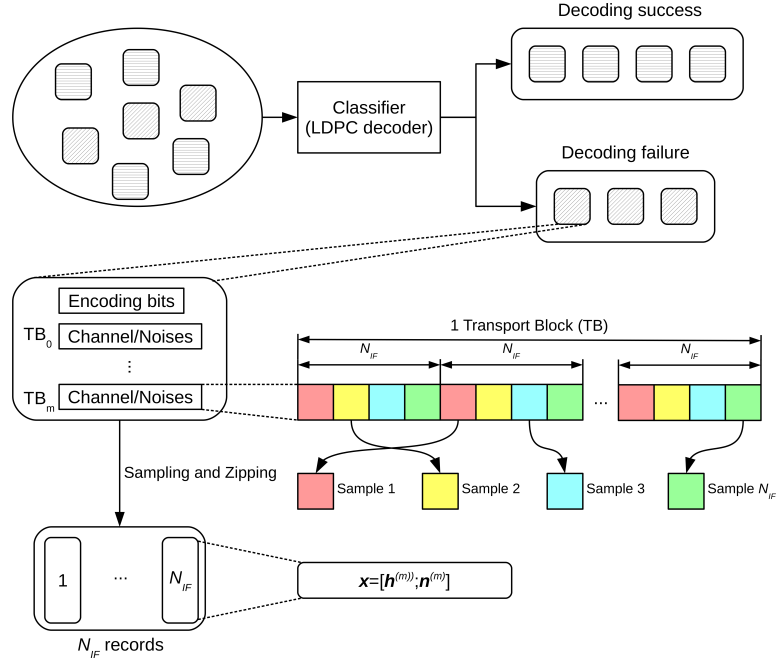}
  \caption{The generation of datasets for the analysis on the posterior
  distribution of the fading channels and noises.}
  \label{fig:model}
\end{figure}

To analyze the posterior distribution of the fading channels and
noises, we generate the channel/noise samples with a LDPC-coded
system~\cite{hochwald2003achieving}\cite{valenti2007coded}. We assume that each
TB contains 1 complete LDPC frame. Another tuning parameter, namely the number of independent fading channels per TB denoted as $N_{IF}$, is added
to this system in order to test the impact of coherence interval on the
posterior distribution. As shown in Fig.~\ref{fig:model}, for different $m$, we
randomly generate a set of LDPC sessions, each consists of a encoding bit sequence and the
fading channel/noise samples corresponding to the $(m+1)$ TBs. The LDPC-decoder
then classify the LDPC sessions into two subsets based on whether the receiver
sends a NACK (decoding failure) or ACK (decoding success) after the $m$-th
retransmission, which represent the two classes of posterior distributions we
are interested in.

Within each TB, the $N_{IF}$ independent fading
channel instances are periodically mapped to the $L_s$ symbols. Among the $L_s$
noise samples, we randomly sample $N_{IF}$ in such a manner that the
corresponding channel samples represent the $N_{IF}$ independent fading
channel instances completely. In this way we make sure that the number of
channel samples and that of the noise samples are equally $N_{IF}$.
Consequently, within each failed/successful HARQ session, a total number of
$(m+1)\times N_{IF}$ groups of fading channels/noises are sampled. This groups of samples are then zipped across the
$(m+1)$ (re)transmissions to construct $N_{IF}$ records, each represented as a
$2(m+1)$-dimensional complex vector, or equivalently a $4(m+1)$-dimensional
real vector, in the form of
\begin{align}
  \mathbf{x} = \left[\Re\{\mathbf{h}^{(m)}\}; \Im\{\mathbf{h}^{(m)}\}; 
  \Re\{\mathbf{n}^{(m)}\}; \Im\{\mathbf{n}^{(m)}\}\right]
\end{align} 
where $\mathbf{h}^{(m)} = [h^{(0)},\ldots,h^{(m)}]$, $\mathbf{n}^{(m)} =
[n^{(0)},\ldots,n^{(m)}]$.

In the next section, we carry out our hypothesis
tests over a dataset of $n$ records of $\mathbf{x}$, which is organized into a $4(m+1)$-by-$n$ matrix
$\mathbf{X} = [\mathbf{x}_1, \ldots, \mathbf{x}_n]$. For notational convenience
we also decompose $X$ into four $(m+1)$-by-$n$ block matrices, i.e.
$\mathbf{X} = [\mathbf{X}_{h,R}; \mathbf{X}_{h,I}; \mathbf{X}_{n, R};
\mathbf{X}_{n, I}]$, which represent the real and imaginary part of the channel
and noise samples, respectively.

\section{Design of Hypothesis Tests}
\label{sec:tests}
In this section, we construct a series of three binary hypothesis tests to see
whether or on what conditions there is a significant difference between the
posterior and the prior joint distribution of the fading channel and noise.
The first test examines whether the posterior data and noise samples follow the same
general type of distributions as the Rician channel and CSCG noise, i.e. whether
$\mathbf{x}$ follows Multi-Variate Normal (MVN) distribution. Once the MVN
distribution is verified, the likelihood of $\mathbf{X}$ can be evaluated,
therefore a second test could further verify whether the distribution of
$\mathbf{x}$ is exactly the same as the prior distribution, i.e. whether the MVN
parameters such as the mean and covariance matches those defined for the prior
distribution. Should the second test fail, we fall back to a third test, which
checks whether the distribution of $\mathbf{x}$ still suggests an i.i.d Rician
fading channel and CSCG noise model, though with potentially different parameters
$\sigma^2$, $K$, $\beta$ and $\theta$ from the prior distribution. If so, the ML estimation of these four parameters could provide some insight into the difference between the
posterior and prior distributions. These three hypothesis tests are detailed as
follows.

\subsection{Test 1: Multi-Variate Normality (MVN) Test}
The first test is literally defined as:
\begin{align}
  \mbox{(Test 1) } \begin{array}{ll}H_0: & \mathbf{x}\mbox{ follows MVN
  distribution.} \\ H_1: & \mbox{Otherwise.} \end{array}
\end{align}
Here we do not make any assumptions on the mean and covariance matrix of
$\mathbf{x}$. As there are a wide variety of MVN tests with different
characteristics~\cite{mecklin2005monte} which may well reach contradictory
conclusions over a same dataset, we adopt the the R package
`MVN'~\cite{korkmaz2014mvn} which implements three popular MVN tests,
namely Mardia's test, Henze-Zirkler(HZ)'s test and Royston's test.

\subsection{Test 2: Parameter Matching Test}
If Test 1 accept the null hypothesis $H_0$, we can
perform a second test to see whether $\mathbf{x}$ has an identical distribution
as the prior assumption. Specifically, assuming that $\mathbf{x}$ is MVN
distributed,
\begin{align}
  \mbox{(Test 2) } \begin{array}{ll}H_0: & \mathbf{x}\sim
  \mathcal{N}(\bm{\mu}_0, \mathbf{\Sigma}_0) \\ H_1: &
  \mbox{Otherwise.}
  \end{array}
\end{align}
where
\begin{subequations}
  \begin{align}
    \bm{\mu}_0 & =
    \left[\sqrt{\frac{K}{K+1}\beta}\cos\theta\mathbf{1}_{(m+1)};
    \sqrt{\frac{K}{K+1}\beta}\sin\theta\mathbf{1}_{(m+1)};
    \mathbf{0}_{2(m+1)}\right]
    \\
    \mathbf{\Sigma}_0 & =
    \mbox{diag}\left(\left[\frac{\beta}{2(K+1)}\mathbf{1}_{2(m+1)},
    \frac{\sigma^2}{2}\mathbf{1}_{2(m+1)}\right]\right) 
  \end{align}.
  \label{eq:mu0Sigma0}
\end{subequations}

The following proposition reduces Test 2 into a chi-squared test:
\begin{proposition}
  \label{prop:test2}
  Test 2 is equivalent to the following hypothesis test:
  \begin{align}
    \mbox{(Test 2(W)) } \begin{array}{ll}H_0: & -2\ln(\Lambda_2) \sim
    \chi_{2(m+1)(4m+7)}^2 \\ H_1: & \mbox{Otherwise.}
    \end{array}
  \end{align}
  as $n\rightarrow\infty$, where
  \begin{align}
    -2\ln(\Lambda_2) & =
    2n(m+1)\ln\left(\frac{\beta\sigma^2}{4(K+1)}\right)
    +
    \frac{2(K+1)(\|\mathbf{X}_{h,R}\|_F^2 + \|\mathbf{X}_{h,I}\|_F^2)}{\beta}
    \notag \\ &+
    \frac{2(\|\mathbf{X}_{n,R}\|_F^2 + \|\mathbf{X}_{n,I}\|_F^2)}{\sigma^2}
    - n\ln|\hat{\mathbf{\Sigma}}| - 4n(m+1)\\
    \hat{\mathbf{\Sigma}} & = \frac{1}{n}(\mathbf{X}
    -\hat{\bm{\mu}}\mathbf{1}_n^T)(\mathbf{X}
    -\hat{\bm{\mu}}\mathbf{1}_n^T)^T,\;\hat{\bm{\mu}} =
    \frac{1}{n}\mathbf{X}\mathbf{1}_n,\label{eq:sigma_mu_ml}
  \end{align}
  in which $\hat{\bm{\Sigma}}$, $\hat{\bm{\mu}}$ are the ML estimation of the
  covariance matrix and mean of $\mathbf{x}$ from $\mathbf{X}$, respectively.
\end{proposition}
\begin{proof}
  See Appendix.
\end{proof}

\subsection{Test 3: Relaxed Parameter Matching Test}
If Test 1 suggests that $\mathbf{x}$ follows MVN
distribution, but Test 2 suggests that it is not identical to
the prior distribution, one would wonder whether the distribution of
$\mathbf{x}$ is still in a manageable form. A natural ``guess'' is that in this
posterior distribution, the channels and noises across the $(m+1)$
(re)transmissions are still i.i.d Rician fading and CSCG distributed,
respectively, except that the parameters $\sigma^2$, $K$, $\beta$ and $\theta$
are different from the prior distribution. In that case, we can examine the
ML estimation of these parameters to see how different they are
from the prior assumption, and whether certain applications are robust against
this parameter mismatch between the posterior and prior distribution. This test
is formulated as follows. Assuming that $\mathbf{x}$ is MVN distributed,
\begin{align}
  \mbox{(Test 3) } \begin{array}{ll}H_0: & \mathbf{x}\sim
  \mathcal{N}(\tilde{\bm{\mu}}_0, \tilde{\mathbf{\Sigma}}_0)
  \\
  H_1:
  &
  \mbox{Otherwise.}
  \end{array}
\end{align}
where $\tilde{\bm{\mu}}_0$ and $\tilde{\mathbf{\Sigma}}_0$ are defined as in
Eq.~\eqref{eq:mu0Sigma0} with $\sigma^2$, $K$, $\beta$, $\theta$ replaced by
unknown parameters $\tilde{\sigma}^2$, $\tilde{K}$, $\tilde{\beta}$ and
$\tilde{\theta}$, respectively.

Similar to Proposition~\ref{prop:test2}, the following proposition reduces Test
3 to a chi-squared test:
\begin{proposition}
  \label{prop:test3}
  Test 3 is equivalent to the following hypothesis test:
  \begin{align}
    \mbox{(Test 3(W)) } \begin{array}{ll}H_0: & -2\ln(\Lambda_3) \sim
    \chi_{2(m+1)(4m+7)-4}^2 \\ H_1: & \mbox{Otherwise.}
    \end{array}
  \end{align}
  as $n\rightarrow\infty$, where
  \begin{align}
    -2\ln(\Lambda_3) & =
    2n(m+1)\ln\left(\frac{\hat{\beta} \hat{\sigma}^2} {4(\hat{K}+1)}\right) -
    n\ln|\hat{\mathbf{\Sigma}}|
  \end{align}
  in which
  \begin{align}
    \hat{\sigma}^2 = \frac{\|\mathbf{X}_{n,R}\|_F^2 +
    \|\mathbf{X}_{n,I}\|_F^2}{n(m+1)} ,\; 
    \hat{K} = \frac{\bar{h}_R^2 + \bar{h}_I^2}{\bar{\sigma}^2_h},\;
    \hat{\beta} = (\hat{K} + 1)\bar{\sigma}^2_h ,\;
    \hat{\theta} =
    \arctan\frac{\bar{h}_I}{\bar{h}_R}, \label{eq:beta_sigma_ml}
  \end{align}
  are the ML estimation of $\tilde{\sigma}^2$, $\tilde{K}$, $\tilde{\beta}$ and
  $\tilde{\theta}$, respectively. The sample mean of the real and imaginary part
  and the sample variance of $h$ are evaluated as
  \begin{subequations}
    \begin{align}
      & \bar{h}_R =
      \frac{\mathbf{1}_{m+1}^T\mathbf{X}_{h,R}\mathbf{1}_n} {n(m+1)} ,\;
      \bar{h}_I = \frac{\mathbf{1}_{m+1}^T\mathbf{X}_{h,I}\mathbf{1}_n}
      {n(m+1)},\\
      & \bar{\sigma}^2_h = \frac{\|\mathbf{X}_{h,R} -
      \bar{h}_R\mathbf{1}_{m+1}\mathbf{1}_{n}^T\|_F^2 +
      \|\mathbf{X}_{h,I} -
      \bar{h}_I\mathbf{1}_{m+1}\mathbf{1}_{n}^T\|_F^2}{n(m+1)}
    \end{align}
  \end{subequations}
\end{proposition}
\begin{proof}
  See Appendix.
\end{proof}

\section{Numerical Results}
\label{sec:numerical}

\subsection{Simulation Settings}
In our simulation, the Rician channel is specified with $\beta = 8$, $K = 1$ and
$\theta = 0$. Each LDPC code word of length $L=2400$ is mapped to 64-QAM
constellation with the same Gray mapping for all (re)transmissions, therefore
each TB consists of $L_s = 400$ symbols. The posterior fading channel and noise
distribution is analyzed at $N_{IF} = 400, 100, 10, 1$ and $m = 0, 1, 2, 3$. For
each pair of $(N_{IF}, m)$, we choose $\sigma^2$ such that around 50\% of HARQ
sessions fail and randomly generate $n\approx 10000$ records of $\mathbf{x}$
for both the failed and successful sessions. % Shall we also consider the
% success
The detailed parameters for each dataset are listed in
Table~\ref{tab:settings}.
\begin{table}[!t]
  \renewcommand{\arraystretch}{1.3}
  \caption{Parameters for each numerically generated dataset.}
  \label{tab:settings}
  \centering
  \begin{tabular}{c|ccccc}
    \hline
    Dataset & $N_{IF}$ & $m$ & $n_{F}$ & $n_{S}$ & $\sigma^2$ \\
    \hline
    1 & 400 & 0 & 10000 & 10000 & 0.13183 \\
    2 & 400 & 1 & 9600  & 10400 & 0.35892 \\
    3 & 400 & 2 & 9600 & 10400 & 0.58479 \\
    4 & 400 & 3 & 9600 & 10400 & 0.84421 \\
    \hline
    5 & 100 & 0 & 10100 & 9900 & 0.13552 \\
    6 & 100 & 1 & 10300 & 9700 & 0.37154 \\
    7 & 100 & 2 & 10100 & 9900 & 0.59772 \\
    8 & 100 & 3 & 10200 & 9800 & 0.83946 \\
    \hline
    9 & 10 & 0 & 10240 & 9760 & 0.13804 \\
    10 & 10 & 1 & 9980 & 10020 & 0.37154 \\
    11 & 10 & 2 & 9970 & 10030 & 0.60534 \\
    12 & 10 & 3 & 10200 & 9800 & 0.84140 \\
    \hline
    13 & 1 & 0 & 10232 & 9768 & 0.18408 \\
    14 & 1 & 1 & 9834 & 10166 & 0.40738 \\
    15 & 1 & 2 & 9915 & 10085 & 0.64565 \\
    16 & 1 & 3 & 9628 & 10372 & 0.86099 \\
    \hline
  \end{tabular}
\end{table}
  
\subsection{Hypothesis Test Results}
The $p$-values of Test 1 over all datasets are shown in Table~\ref{tab:test1}.
For $N_{IF} = 1$, the posterior distribution indeed appears to be of different
type from the prior distribution. However, for $N_{IF} = 10, 100, 400$, there is
strong evidence indicating that the fading channels and noises still follow MVN
distribution. This is especially true for larger $N_{IF}$ (100, 400) where all
the three MVN tests support the null hypothesis, and the failed sessions
in which we are more interested since they are supposed to be used in the
posterior analysis.

The $p$-values of Test 2 and Test 3 are shown in Table~\ref{tab:test23}. As we can see, for larger $N_{IF}$
 (100, 400), Test 2 suggests that the posterior distribution are likely to be
the same as the prior distribution. For intermediate $N_{IF} = 10$, although
Test 2 rejects the null hypothesis, in 3 of the 4 failed sessions Test 3
indicates that the posterior distribution can be still viewed as independent Rician fading
channels and CSCG noises. A closer look at the ML estimation results
in Table~\ref{tab:ML} reveals that the posterior $\hat{\beta}$ and
$\hat{\theta}$ are almost the same as the prior ones. However, the posterior
Rician channel has smaller $\hat{\beta}$ and $\hat{K}$ for the failed sessions,
and larger ones for the successful sessions, as compared to the prior
parameters.
In general, this gap increases as $N_{IF}$ and $m$ decreases. Finally, when
$N_{IF} = 1$, despite the rejection of MVN hypothesis so that Test 2 and Test 3
are meaningless, the ML estimation in Eq.~\eqref{eq:beta_sigma_ml} still serves
as the empirical mean square of the fading channel and the noises. The noise
power is still almost the same as the prior distribution, while the channel
power is much smaller and much larger for the failed and successful
sessions, respectively, than the prior one.

\begin{table}[!t]
  \renewcommand{\arraystretch}{1.3}
  \caption{Test 1: MVN tests. The $p$-values
  correponding to rejected null hypotheses ($p < 0.01$) are colored red.}
  \label{tab:test1}
  \centering
  \begin{tabular}{c|cc|c|c|cc|c|c}
    \hline
    \multirow{2}{*}{Dataset}  & \multicolumn{4}{c|}{Failed sessions} &
    \multicolumn{4}{c}{Successful sessions} \\
    \cline{2-9}
    & Mardia-skew & Mardia-kurt & HZ & Royston & Mardia-skew & Mardia-kurt & HZ
    & Royston \\
    \hline
    1 & 0.03707 & 0.5945 & 0.1044 &
    0.09758 & 0.3086 & 0.3285 & 0.1810 &
    0.5405
    \\
    2 & 0.2869 & 0.6188 & 0.9854 &
    0.2496 & 0.7373 & 0.7304 & 0.8650 &
    0.2439
    \\
    3 & 0.7019 & 0.05709 & 0.9187 &
    0.4361 & 0.7523 & 0.1693 & 0.7661 &
    0.1821
    \\
    4 & 0.7590 & 0.7925 & 0.5395 &
    0.2098 & 0.6935 & 0.4394 & 0.9345 & 0.8580
    \\
    \hline
    5 & 0.2720 & 0.1313 & 0.2117 &
    0.9635 & 0.6373 & 0.7678 & 0.1052 & 0.1444
    \\
    6 & 0.8645 & 0.2425 & 0.4166 &
    0.7259 & 0.5645 & 0.7191 & 0.1495 &
    0.9247 \\
    7 & 0.6248 & 0.9021 & 0.9473 &
    0.4020 & 0.1583 & 0.7446 & 0.4879 &
    0.8501
    \\
    8 & 0.1308 & 0.7417 & 0.8127 &
    0.1612 & 0.9160 & 0.5376 & 0.3666 &
    0.3905 \\
    \hline
    9 & \tred{6.172e-10} & 0.02679 & \tred{8.639e-13} &
    0.01571 & \tred{7.459e-20} & 0.04260 & \tred{6.661e-16} & \tred{2.562e-4} \\
    10 & \tred{9.934e-4} & 0.3654 & 0.01845 &
    0.02896 & \tred{7.720e-8} & 0.8683 & \tred{2.494e-4} & 0.01850 \\
    11 & 0.3692 & 0.03412 & 0.4131 &
    0.7423 & 0.1039 & 0.1747 & 0.5589 &
    0.8595 \\
    12 & 0.6141 & 0.9034 & 0.7274 &
    0.04015 & 0.8693 & 0.1940 & 0.6021 &
    0.4035
    \\
    \hline
    13 & \tred{1.414e-229} & \tred{9.191e-6} & \tred{0} & \tred{2.753e-27} &
    \tred{0} & \tred{0} & \tred{0} & \tred{1.646e-22} \\
    14 & \tred{5.118e-236} & \tred{4.239e-8} & \tred{0} & \tred{2.613e-22} &
    \tred{5.620e-269} & \tred{1.800e-7} & \tred{0} & \tred{3.049e-12}
    \\
    15 & \tred{5.052e-156} & \tred{6.174e-10} & \tred{0} & \tred{2.596e-14} &
    \tred{3.303e-243} & 0.01597 & \tred{0} & \tred{3.497e-10}
    \\
    16 & \tred{3.235e-131} & \tred{3.478e-10} & \tred{0} & \tred{2.356e-13} &
    \tred{2.006e-173} & 0.9775 & \tred{0} & \tred{1.130e-6} \\
    \hline
  \end{tabular}
\end{table}
  
\begin{table}[!t]
  \renewcommand{\arraystretch}{1.3}
  \caption{Test 2 and Test 3, parameter matching tests (exact and relaxed).
  The $p$-values correponding to rejected null hypotheses ($p < 0.01$) are
  colored red.}
  \label{tab:test23}
  \centering
  \begin{tabular}{c|c|c|c|c}
    \hline
    \multirow{2}{*}{Dataset} &  \multicolumn{2}{c|}{Failed sessions} &
    \multicolumn{2}{c}{Successful sessions} \\
    \cline{2-5}
    & Test 2 & Test 3 & Test 2 & Test 3 \\
    \hline
    1 & 0.1350 & 0.2781 & 0.04330 & 0.3123 \\
    2 & 0.1614 & 0.5014 & 0.9206 & 0.9258 \\
    3 & 0.1125 & 0.1436 & 0.1333 & 0.2081 \\
    4 & 0.4806 & 0.4941 & 0.9758 & 0.9734 \\
    \hline
    5 & 0.01253 & 0.04575 & \tred{1.358e-11} & \tred{7.713e-3} \\
    6 & 0.06204 & 0.3005 & 0.02778 & 0.6205 \\
    7 & 0.2207 & 0.6941 & 0.3504 & 0.7882 \\
    8 & 0.4434 & 0.9241 & 0.01312 & 0.2772 \\
    \hline
    9 & \tred{0} & \tred{5.236e-4} & \tred{0} & \tred{6.890e-10} \\
    10 & \tred{0} & 0.02033 & \tred{0} & \tred{1.264e-5} \\
    11 & \tred{0} & 0.3848 & \tred{0} & \tred{6.695e-8} \\
    12 & \tred{0} & 0.3573 & \tred{0} & \tred{6.697e-3} \\
    \hline
    13 & \tred{0} & \tred{0} & \tred{0} & \tred{0} \\
    14 & \tred{0} & \tred{0} & \tred{0} & \tred{0} \\
    15 & \tred{0} & \tred{0} & \tred{0} & \tred{0} \\
    16 & \tred{0} & \tred{0} & \tred{0} & \tred{0} \\
    \hline
  \end{tabular}
\end{table}

\begin{table}[!t]
  \renewcommand{\arraystretch}{1.3}
  \caption{The ML estimation of the parameters of Rician channel and CSCG
  noise from Test 3.}
  \label{tab:ML}
  \centering
  \begin{tabular}{c|cccc|cccc}
    \hline
    \multirow{2}{*}{Dataset}  & \multicolumn{4}{c|}{Failed sessions} &
    \multicolumn{4}{c}{Successful sessions} \\
    \cline{2-9}
    & $\hat{\beta}$ & $\hat{K}$ & $\hat{\theta}$ & $\hat{\sigma}^2$
    & $\hat{\beta}$ & $\hat{K}$ & $\hat{\theta}$ & $\hat{\sigma}^2$ \\
    \hline
    1 & 7.830 & 0.9883 & 6.391e-3 &
    0.1330 & 8.137 & 1.015 & -9.221e-3 & 0.1284
    \\
    2 & 7.902 & 0.9899 & -6.634e-3 &
    0.3664 & 8.063 & 1.017 & -2.772e-3 & 0.3572 \\
    3 & 7.936 & 0.9920 & 4.855e-3 &
    0.5904 & 8.027 & 1.010 & -8.987e-3 & 0.5797
    \\
    4 & 8.017 & 0.9954 & -6.380e-3 &
    0.8479 & 8.011 & 1.013 & -1.333e-3 & 0.8468 \\
    \hline
    5 & 7.831 & 0.9793 & 1.878e-3 &
    0.1381 & 8.505 & 1.078 & -3.657e-3 & 0.1336
    \\
    6 & 7.827 & 0.9952 & -4.889e-3
    & 0.3733  & 8.232 & 1.017 & -0.01006 & 0.3683
    \\
    7 & 7.854 & 0.9692 & -3.902e-3 &
    0.6032 & 8.158 & 1.022 & -7.301e-3 & 0.5987
    \\
    8 & 7.822 & 0.9765 & -3.594e-3 &
    0.8413 & 8.194 & 1.035 & 2.162e-4 & 0.8369 \\
    \hline
    9 & 6.760 & 0.8111 & -3.474e-3 &
    0.1418 & 9.262 & 1.261 & 4.441e-3 & 0.1373
    \\
    10 & 6.965 & 0.8721 & -3.151e-3 &
    0.3726 & 9.046 & 1.142 & 6.379e-3 & 0.3740
    \\
    11 & 7.105 & 0.8999 & -2.120e-3 &
    0.6079 & 8.936 & 1.118 & 1.622e-3 & 0.6039 \\
    12 & 7.240 & 0.9198 & 4.991e-4 &
    0.8442 & 8.758 & 1.078 & 6.637e-3 & 0.8363 \\
    \hline
    13 & 2.958 & 0.6028 & 0.01418 &
    0.1860 & 13.20 & 2.101 & -5.829e-3 & 0.1876
    \\
    14 & 4.206 & 0.7147 & -5.320e-3 &
    0.4027 & 11.67 & 1.522 & -2.713e-3 & 0.4038 \\
    15 & 4.911 & 0.7668 & -3.975e-4 &
    0.6421 & 11.06 & 1.381 & 2.002e-3 & 0.6428 \\
    16 & 5.247 & 0.7996 & 1.944e-3 &
    0.8598 & 10.56 & 1.305 & -2.141e-3 & 0.8551 \\
    \hline
  \end{tabular}
\end{table}
  
\subsection{The Impact of Parameter Mismatch on MoDiv Design}
\begin{figure}[htb]
  \begin{minipage}[b]{.48\linewidth}
    \centering
    \centerline{\includegraphics[width=7cm]{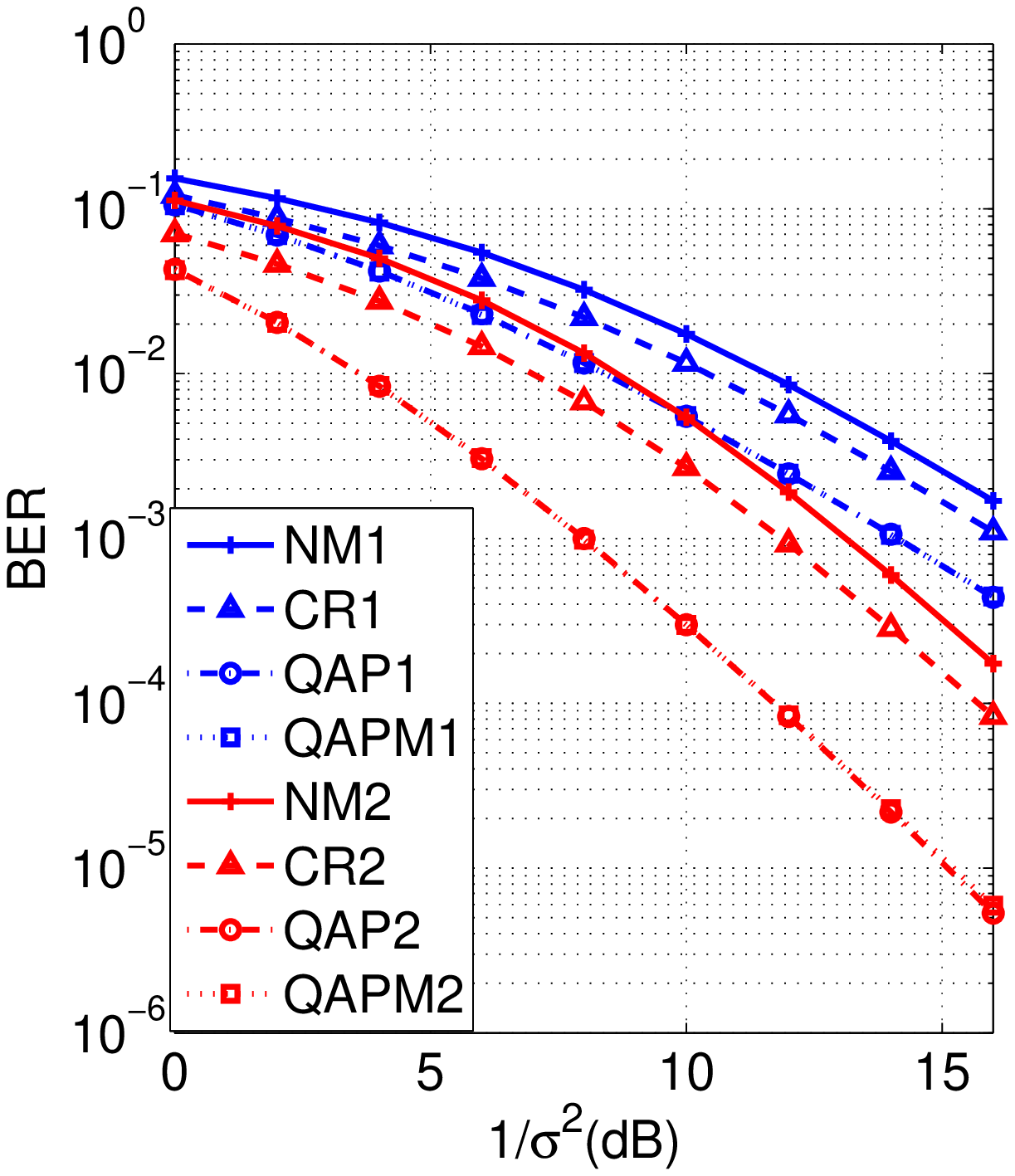}}
    \centerline{(a) $m=1,2$}\medskip
  \end{minipage}
  \hfill
  \begin{minipage}[b]{0.48\linewidth}
    \centering
    \centerline{\includegraphics[width=7cm]{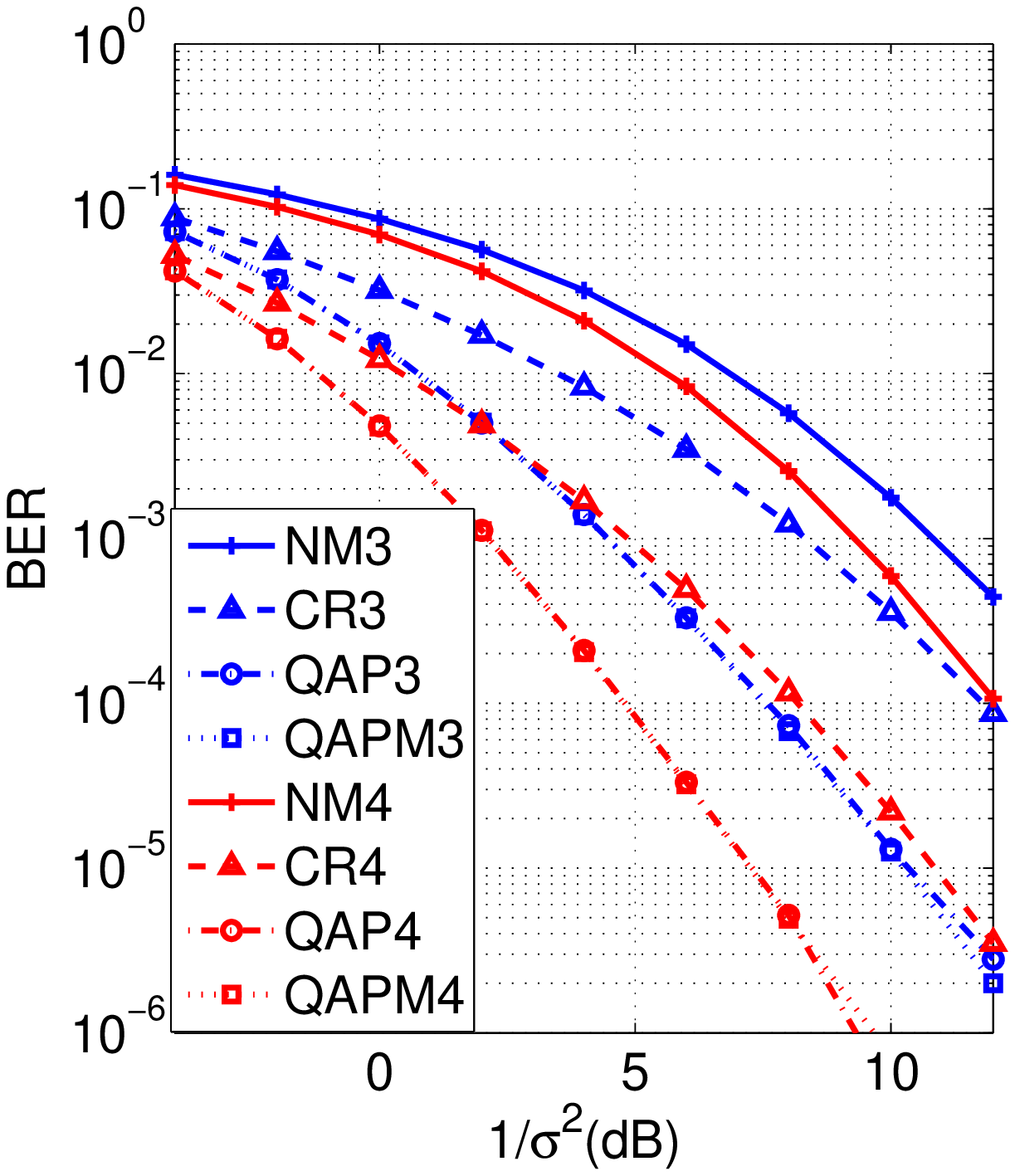}}
    \centerline{(b) $m=3,4$}\medskip
  \end{minipage}
  \caption{The Monte-Carlo simulated uncoded BER for different $m$. NM$m$ and
  CR$m$ are two benchmark modulation diversity schemes. QAPM$m$ represents the
  QAP-based MoDiv scheme designed using prior distribution but tested on the
  posterior distribution, while QAP$m$ represents the QAP-based MoDiv scheme
  designed and tested on the posterior distribution.}
  \label{fig:mismatch}
\end{figure}

For the intermediate $N_{IF} = 10$, the hypothesis tests results suggest that
the difference between the posterior and prior distribution can be viewed
mostly as a parameter mismatch and the main difference is on the channel power
$\beta$ and Rician factor $K$. As an example of how this parameter will affect
certain type of application, we consider the Modulation Diversity (MoDiv) design
problem in~\cite{wu2015modulation}\cite{harvind2005symbol}. we compare the
MoDiv design based on prior distribution ($\beta = 8$ and $K = 1$) and that
based on an artificial posterior distribution for the failed sessions ($\beta =
6.5$ and $K = 0.8$), and compare their actual BER performance over the
artificial posterior distribution.
The simulation results are shown in Fig.~\ref{fig:mismatch}.
Despite that the gap in $K$ and $\beta$ between the two distributions is larger
than suggested by Table~\ref{tab:test23}, there is hardly any difference in BER
performance, suggesting that MoDiv design is an application somehow robust
against the difference between the posterior and prior distribution.

\subsection{Remarks}
These numerical results demonstrate that, except when $N_{IF}$ is very small,
the posterior distribution conditioned on the failure of all previous HARQ
transmissions is not so different from the prior distribution, or this
difference is too trivial to affect certain type of applications. 
In many existing works on HARQ, no assumption on $N_{IF}$ or the coherence
interval is made except that the channels are independent across
retransmissions, and no FEC/CRC scheme or transport block structure is
specified. Attempt to use the posterior distribution is likely to be unfounded.
On the other hand, when we introduce more general, practical assumptions to the
HARQ systems being studied, it is usually difficult, if not impossible, to
derive the posterior distribution.

\section{Conclusion}
\label{sec:conclusion}
This work investigates the posterior fading channel and noise distribution in
a HARQ system conditioned on the HARQ feedbacks. We design three hypothesis tests,
demonstrating with channel and noise samples generated from a practical
LDPC-coded HARQ system that, unless the coherence interval is large or the
number of independent fading instances per TB is small, the posterior distribution is
not so different from the prior distribution, or the difference is so small that
it has negligible effect on certain types of applications. To some extend, this
work justifies the seemingly lax use of prior distribution in many existing
works about HARQ.

\appendix[Proof for Proposition~\ref{prop:test2} and
Proposition~\ref{prop:test3}]
\label{append:proof}
The proof is simply an adoption of Wilks' theorem~\cite{wilks1938large}. The
log-likelihood of observing $\mathbf{X}$ under MVN distribution
$\mathcal{N}(\bm{\mu}, \mathbf{\Sigma})$ is
\begin{align}
  \ln(L(\bm{\mu}, \mathbf{\Sigma}|\mathbf{X})) =
  -\frac{n}{2}\ln|\mathbf{\Sigma}|-
  \frac{1}{2}\mbox{tr}(\mathbf{\Sigma}^{-1}(\mathbf{X}
  -\bm{\mu}\mathbf{1}_n^H)(\mathbf{X} -\bm{\mu}\mathbf{1}_n^H)^H) + C  
\end{align}
where $C$ is a constant. The parameter space $\Theta = \{(\bm{\mu},
\mathbf{\Sigma})\}$ has a dimension of $2(m+1)(4m+7)$. Its supremum
\begin{align}
  \sup \{\ln(L(\bm{\mu}, \mathbf{\Sigma}|\mathbf{X}))\} & =
  \ln(L(\hat{\bm{\mu}}, \hat{\mathbf{\Sigma}}|\mathbf{X})) \notag\\
  & = -\frac{n}{2}\ln|\hat{\mathbf{\Sigma}}| - 2n(m+1) + C
\end{align}
where $\hat{\bm{\mu}}, \hat{\mathbf{\Sigma}}$ are defined in
Eq.~(\ref{eq:sigma_mu_ml}).

Under the null hypothesis of Test 2, the parameter space $\Theta_0^{(T2)}$ has 0
dimensionality, the supremum of log-likelihood is
\begin{align}
  & \sup \{\ln(L(\bm{\mu}, \mathbf{\Sigma}|\mathbf{X})):\,(\bm{\mu},
  \mathbf{\Sigma})\in \Theta_0^{(T2)}\} \notag \\
  = & \ln(L(\bm{\mu}_0,
  \mathbf{\Sigma}_0|\mathbf{X})) \notag \\
  = & -n(m+1)
  \ln\left(\frac{\beta \sigma^2}{4(K+1)}\right) -
  \frac{(K+1)(\|\mathbf{X}_{h,R}\|_F^2 + \|\mathbf{X}_{h,I}\|_F^2)}{\beta} -
  \frac{\|\mathbf{X}_{n,R}\|_F^2 + \|\mathbf{X}_{n,I}\|_F^2}{\sigma^2}
  + C
\end{align}
therefore according to Wilks' theorem, as $n\rightarrow\infty$ 
\begin{align}
  -2\ln(\Lambda_2) &= 2\left(\sup \{\ln(L(\bm{\mu},
  \mathbf{\Sigma}|\mathbf{X}))\} - \sup \{\ln(L(\bm{\mu}, \mathbf{\Sigma}|\mathbf{X})):\,(\bm{\mu},
  \mathbf{\Sigma})\in \Theta_0^{(T2)}\}\right) \notag \\
  &\sim \chi_{2(m+1)(4m+7)}^2
\end{align}

Under the null hypothesis of Test 3, the parameter space $\Theta_0^{(T3)}$
has a dimensionality of 4, the supremum of log-likelihood is
\begin{align}
  & \sup \{\ln(L(\bm{\mu}, \mathbf{\Sigma}|\mathbf{X})):\,(\bm{\mu},
  \mathbf{\Sigma})\in \Theta_0^{(T3)}\} \notag \\
  = & \ln(L(\hat{\bm{\mu}}_0,
  \hat{\mathbf{\Sigma}}_0|\mathbf{X})) \notag \\
  = & -n(m+1)\ln\left(\frac{\hat{\beta} \hat{\sigma}^2} {4(\hat{K}+1)}\right) -
    2n(m+1) + C
\end{align}
where $\hat{\bm{\mu}}_0 $, $\hat{\mathbf{\Sigma}}_0 $ are defined as in
Eq.~\eqref{eq:mu0Sigma0} with $\sigma^2$, $K$, $\beta$, $\theta$ replaced by
the ML estimation $\hat{\sigma}^2$, $\hat{K}$,
$\hat{\beta}$, $\hat{\theta}$, respectively. Similar to the case of Test 2,
as $n\rightarrow\infty$
\begin{align}
  -2\ln(\Lambda_3) &= 2\left(\sup \{\ln(L(\bm{\mu},
  \mathbf{\Sigma}|\mathbf{X}))\} - \sup \{\ln(L(\bm{\mu}, \mathbf{\Sigma}|\mathbf{X})):\,(\bm{\mu},
  \mathbf{\Sigma})\in \Theta_0^{(T3)}\}\right) \notag \\
  &\sim \chi_{2(m+1)(4m+7) - 4}^2
\end{align}
\bibliographystyle{IEEEtran}
% argument is your BibTeX string definitions and bibliography database(s)
\bibliography{IEEEabrv,./refs.bib}

% Generated by IEEEtran.bst, version: 1.12 (2007/01/11)
\begin{thebibliography}{10}
\providecommand{\url}[1]{#1}
\csname url@samestyle\endcsname
\providecommand{\newblock}{\relax}
\providecommand{\bibinfo}[2]{#2}
\providecommand{\BIBentrySTDinterwordspacing}{\spaceskip=0pt\relax}
\providecommand{\BIBentryALTinterwordstretchfactor}{4}
\providecommand{\BIBentryALTinterwordspacing}{\spaceskip=\fontdimen2\font plus
\BIBentryALTinterwordstretchfactor\fontdimen3\font minus
  \fontdimen4\font\relax}
\providecommand{\BIBforeignlanguage}[2]{{%
\expandafter\ifx\csname l@#1\endcsname\relax
\typeout{** WARNING: IEEEtran.bst: No hyphenation pattern has been}%
\typeout{** loaded for the language `#1'. Using the pattern for}%
\typeout{** the default language instead.}%
\else
\language=\csname l@#1\endcsname
\fi
#2}}
\providecommand{\BIBdecl}{\relax}
\BIBdecl

\bibitem{wu2015modulation}
W.~Wu, H.~Mittelmann, and Z.~Ding, ``{Modulation design for two-way
  amplify-and-forward relay HARQ},'' submitted for publication.

\bibitem{gu2006modeling}
J.~Gu, Y.~Zhang, and D.~Yang, ``{Modeling conditional FER for hybrid ARQ},''
  \emph{{IEEE} Commun. Lett.}, vol.~10, no.~5, pp. 384--386, May 2006.

\bibitem{long2012analysis}
H.~Long, W.~Xiang, S.~Shen, Y.~Zhang, K.~Zheng, and W.~Wang, ``{Analysis of
  conditional error rate and combining schemes in HARQ},'' \emph{{IEEE} Trans.
  Signal Process.}, vol.~60, no.~5, pp. 2677--2682, May 2012.

\bibitem{alkurd2015modeling}
R.~Alkurd, R.~Shubair, and I.~Abualhaol, ``Modeling conditional error
  probability for hybrid decode-amplify-forward cooperative system,'' in
  \emph{Proc. IEEE Wireless Commun. Netw. Conf. (WCNC)}, March 2015, pp. 7--12.

\bibitem{harvind2005symbol}
H.~Samra, Z.~Ding, and P.~Hahn, ``Symbol mapping diversity design for multiple
  packet transmissions,'' \emph{{IEEE} Trans. Commun.}, vol.~53, no.~5, pp.
  810--817, May 2005.

\bibitem{chaitanya2014adaptive}
T.~Chaitanya and E.~Larsson, ``Adaptive power allocation for {HARQ} with
  {Chase} combining in correlated rayleigh fading channels,'' \emph{IEEE
  Wireless Commun. Lett.}, vol.~3, no.~2, pp. 169--172, April 2014.

\bibitem{jin2011optimal}
H.~Jin, C.~Cho, N.-O. Song, and D.~K. Sung, ``Optimal rate selection for
  persistent scheduling with harq in time-correlated nakagami-m fading
  channels,'' \emph{IEEE Trans. Wireless Commun.}, vol.~10, no.~2, pp.
  637--647, February 2011.

\bibitem{ts36.213}
{3GPP TS36.213}, ``{Evolved Universal Terrestrial Radio Access (E-UTRA):
  Physical layer procedures},'' Sep. 2015, v12.7.0.

\bibitem{ts36.141}
{3GPP TS36.141}, ``{Evolved Universal Terrestrial Radio Access (E-UTRA): Base
  Station (BS) conformance testing},'' Oct. 2015, v13.1.0.

\bibitem{tse2005fundamentals}
D.~Tse and P.~Viswanath, \emph{Fundamentals of wireless communication}.\hskip
  1em plus 0.5em minus 0.4em\relax Cambridge university press, 2005.

\bibitem{hochwald2003achieving}
B.~Hochwald and S.~ten Brink, ``Achieving near-capacity on a multiple-antenna
  channel,'' \emph{{IEEE} Trans. Commun.}, vol.~51, no.~3, pp. 389--399, March
  2003.

\bibitem{valenti2007coded}
\BIBentryALTinterwordspacing
M.~Valenti. (2007) The coded modulation library. [Online]. Available:
  \url{http://www.iterativesolutions.com}
\BIBentrySTDinterwordspacing

\bibitem{mecklin2005monte}
C.~J. Mecklin and D.~J. Mundfrom, ``A {Monte Carlo} comparison of the {Type I
  and Type II} error rates of tests of multivariate normality,'' \emph{Journal
  of Statistical Computation and Simulation}, vol.~75, no.~2, pp. 93--107,
  2005.

\bibitem{korkmaz2014mvn}
S.~Korkmaz, D.~Goksuluk, and G.~Zararsiz, ``Mvn: An r package for assessing
  multivariate normality,'' \emph{A peer-reviewed, open-access publication of
  the R Foundation for Statistical Computing}, p. 151, 2014.

\bibitem{wilks1938large}
S.~S. Wilks, ``The large-sample distribution of the likelihood ratio for
  testing composite hypotheses,'' \emph{The Annals of Mathematical Statistics},
  vol.~9, no.~1, pp. 60--62, 1938.

\end{thebibliography}

% that's all folks
\end{document}